\documentclass[aps,twocolumn,superscriptaddress,longbibliography]{revtex4-2} 
\usepackage{amssymb,amsmath,amsthm,amsfonts,amsbsy,mathrsfs}
\usepackage{graphicx}
\usepackage{color}
\usepackage{bm}

\usepackage[%
  colorlinks=true,
  urlcolor=blue,
  linkcolor=blue,
  citecolor=blue
]{hyperref}

\newcommand{\oL}{\omega_{\rm L}}
\newcommand{\oH}{\omega_{\rm H}}
\newcommand{\oD}{\omega_{\rm D}}
\newcommand{\lD}{\lambda_{\rm D}}
\newcommand{\xib}{\xi_{\rm bulk}}
\newcommand{\red}[1]{\textcolor{red}{#1}}

\begin{document}
\title{Long-wavelength fluctuations and dimensionality crossover in confined liquids}
\author{Jing Yang}
\affiliation{Division of Physics and Applied Physics, School of Physical and
Mathematical Sciences, Nanyang Technological University, Singapore 637371, Singapore}
\author{Yan-Wei Li}
\affiliation{Division of Physics and Applied Physics, School of Physical and
Mathematical Sciences, Nanyang Technological University, Singapore 637371, Singapore}
\author{Massimo Pica Ciamarra}
\email{massimo@ntu.edu.sg}
\affiliation{Division of Physics and Applied Physics, School of Physical and
Mathematical Sciences, Nanyang Technological University, Singapore 637371, Singapore}
\affiliation{
CNR--SPIN, Dipartimento di Scienze Fisiche,
Universit\`a di Napoli Federico II, I-80126, Napoli, Italy
}
\date{\today}

\begin{abstract}
The phase behavior of liquids confined in a slit geometry does not reveal a crossover from a three- to a two-dimensional behavior as the gap size decreases.
Indeed, the prototypical two-dimensional hexatic phase only occurs in liquids confined to a monolayer. 
Here, we demonstrate that the dimensionality crossover is apparent in the lateral size dependence of the relaxation dynamics of confined liquids, developing a Debye model for the density of vibrational states of confined systems and performing extensive numerical simulations.
In confined systems, Mermin-Wagner fluctuations enhance the amplitude of vibrational motion or Debye-Waller factor by a quantity scaling as the inverse gap width and proportional to the logarithm of the aspect ratio, as a clear signature of a two-dimensional behaviour.
As the temperature or lateral system size increases, the crossover to a size-independent relaxation dynamics occurs when structural relaxation takes place before the vibrational modes with the longest wavelength develop\red{.}
\end{abstract}
\maketitle

\section{Introduction}
The phase behaviour and dynamics of liquids confined in slit geometries are affected by the competition of several length scales.
Indeed, for a liquid confined in a slit of dimension $L \times L \times H$, the lateral length $L$ and gap width $H \ll L$ interfere with bulk-liquid length scales, such as the typical distance between the particles, $a_0 = \rho^{-1/3}$, and the structural correlation length, $\xib \simeq 10$, e.g., as estimated from the decay of the radial distribution function~\cite{Granick1991, Mandal2014, Zhang2020}.
The competition between $H$ and $\xib$ induces a cascade of confinement-induced ordering transitions~\cite{Schmidt1996, Lowen2009, Cummings2010, Mandal2014}, and a solid like behaviour interpreted as a signal of a first-order transition~\cite{Klein1995, Klein1998} or, more recently~\cite{Demirel2001, Zhu2004, Kienle2016}, as a continuous glass transition. 
For molecular liquids in very narrow confinements, length scales associated with the anisotropic molecular structure~\cite{Granick1991,jabbarzadeh2016friction,jabbarzadeh2006low,jabbarzadeh2007crystal} and the details of the interaction between the molecules and the confining walls also play a role.

The rich and system-dependent phase behaviour of confined systems makes difficult rationalizing the crossover from three to two dimensions focusing on its gap size dependence.
Indeed the hexatic phase, which is a phase with short-ranged translational order and long-ranged bond-orientational order only occurring in two-dimensional systems, has been only reported for $H \simeq a_0$ in Lennard-Jones systems~\cite{Gribova2011}. 
In this extremely confined limit, the occurrence of a two-dimensional behaviour is in line with the observed decoupling of the lateral and transverse degrees of freedom~\cite{Franosch2012, Mandal2017}.

The size dependence of the relaxation dynamics of confined liquids offers an alternative and unexplored approach to investigate the dimensionality crossover.
Indeed, two-dimensional systems differ from their three-dimensional counterpart because Mermin-Wagner~\cite{mermin1966absence} long-wavelength (LW) fluctuations make their relaxation dynamics size dependent~\cite{flenner2015fundamental, shiba2016unveiling, illing2017mermin, vivek2017long, zhang2019long, shiba2019local, li2019long}. 
This alternative approach is also convenient as Mermin-Wagner fluctuations are always present in two-dimensional systems; conversely, the two- and the three-dimensional phase behaviour do not qualitatively differ in all systems~\cite{Krauth2011, Anderson2017, Li2020}.

In this paper, we demonstrate that confined systems have a relaxation dynamics depending on the lateral size $L$, as two-dimensional ones, and rationalize the dimensionality crossover clarifying how this $L$ dependence varies with the gap width $H$ and relaxation time.
We find that, in the solid regime, confinement {\it enhances} the asymptotic value of the mean-square displacement, or Debye-Waller factor, by a factor scaling as $(1/H)\ln(L/H)$.
A similar enhancement of the mean square displacement occurs in the liquid phase. 
Liquids, however, exhibit a dimensionality crossover as size-effects vanish above a characteristic $H$-independent system size fixed by sound velocity and relaxation time.
We further clarify that our predictions apply to both molecular and colloidal liquids through the investigation of experimentally relevant confinement settings.

\section{Debye's DOS in confinement} 
\begin{figure*}[!t]
\centering
\includegraphics[angle=0,width=2\columnwidth]{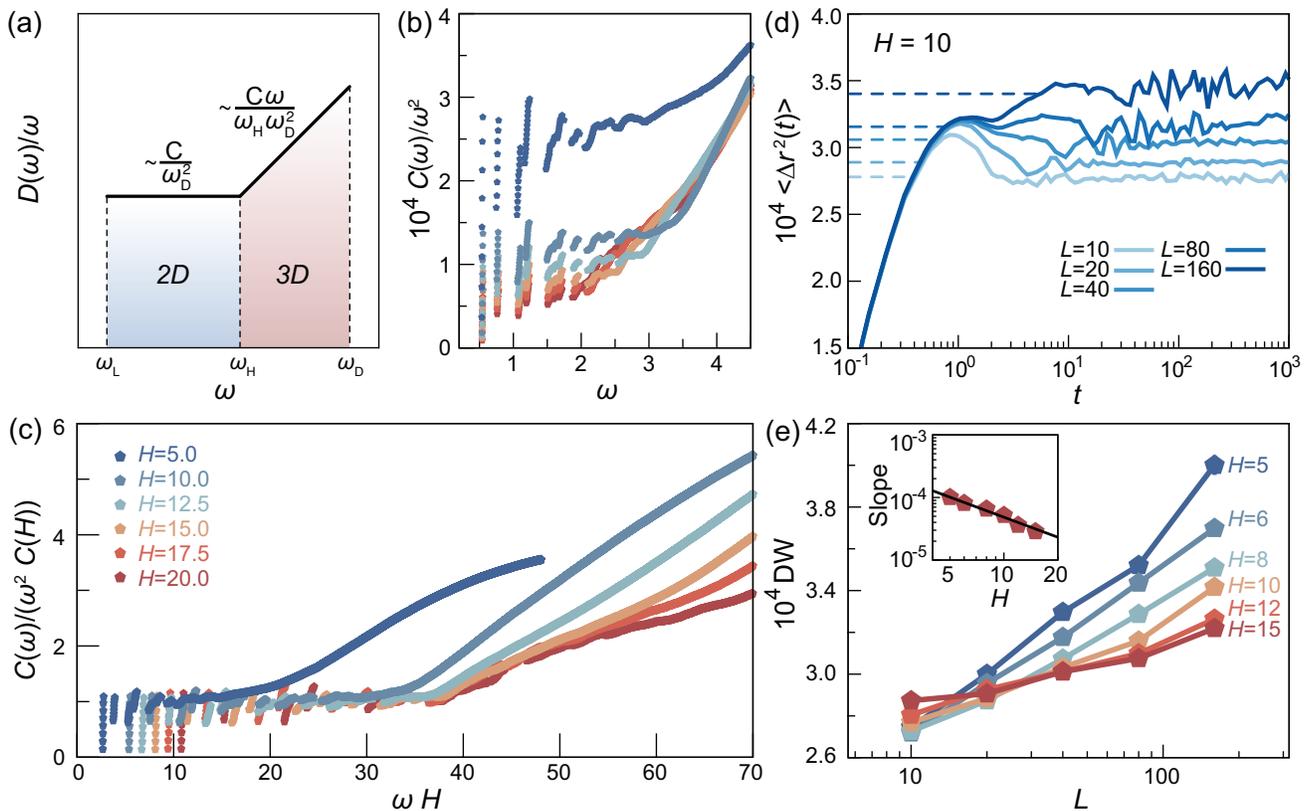}
\caption{
(a) Schematic illustration of the Debye's density of states of quasi-2D systems, Eq.~\ref{eq:Domega}. (b) Low-frequency cumulative density of states of confined solids with lateral length $L=80$ and different gap sizes $H$. (c) The data in ${\bf a}$ collapses when plotted vs. $\omega H$ and vertically scaled, for $H > \xib \simeq 10$. (d) Mean square displacement at $T = 0.005$ and $H = 10$, for different $L$ values. (e) The asymptotic DW factor grows logarithmically with the lateral size $L$, with a slope scaling as $1/H$ (inset). Errors are smaller than the symbol size.
\label{fig:dos}
}
\end{figure*}

We develop a Debye-like model for the vibrational density of states (DOS) of confined amorphous solids to rationalize the size dependence of their dynamical properties.
In confinement, the length scales $L$ and $H$ and the transverse sound velocity $c_s$ fix two characteristic frequencies, $\oL = 2\pi c_s/L$ and $\oH = 2\pi c_s/H$.
$\omega_{\rm L}$ is the smallest possible phonon frequency. 
The physical role of $\oH$ is understood considering that phonons with $\omega < \oH$, which have a wavelength larger than $H$, do not fit along the transverse direction.
Hence $\oH$ separates the spectrum into a low-frequency region, $\oL < \omega < \oH$ where excitations are essentially two dimensional, and in a high frequency region, $\oH < \omega < \oD$, with $\oD$ the Debye' frequency, where excitations are three dimensional.
In the Debye' approximation, the density of states is
\begin{equation}
    D(\omega) = \left\{
\begin{aligned}
    &c~\frac{\omega}{\oD^2} & & {\oL \leq \omega \leq \oH} \\
    &c~ \frac{\omega^2}{\oH\oD^2} & & {\oH \leq \omega \leq \oD},
\end{aligned}
\right.\label{eq:Domega}
\end{equation}
with $c$ non-dimensional normalization constant, 
\begin{equation}
c^{-1} = \frac{1}{2} \left[\left(\frac{\oH}{\oD}\right)^2-\left(\frac{\oL}{\oD}\right)^2\right]-\frac{1}{3}\left[\frac{\oD}{\oH}-\left(\frac{\oH}{\oD}\right)^3\right].
\end{equation}
$D(\omega)$ is schematically illustrated in Fig.
\ref{fig:dos}(a).
We remark that we have restricted the above investigation to the transverse modes, which are of greater relevance to our purposes as having a smaller frequency. 
The longitudinal modes can be similarly described.

The vibrational density of states allows us to evaluate the asymptotic value of the mean square displacement, or the Debye-Waller factor, averaging the contributions $k_B T/m\omega^2$ of the different modes.
To highlight the dependence on the different length scales involved, we write $\oD = 2\pi c_s/\lD$, finding
\begin{equation}
    {\rm DW}= \frac{k_BT}{m\oD^2} \frac{ \ln \left(\frac{L}{H}\right)+\frac{H}{\lD}-1}{\frac{1}{2}\left[ \left(\frac{\lD}{H}\right)^2-\left(\frac{\lD}{L}\right)^2 \right]+\frac{1}{3}\left[\frac{H}{\lD}-\left(\frac{\lD}{H}\right)^2\right]}.
    \label{eq:dw}
\end{equation}
The three dimensional limit, ${\rm DW_{3D}}\simeq \frac{3 k_BT}{m\oD^2}$, and the two-dimensional one, ${\rm DW_{2D}}=\frac{2 k_BT}{m\oD^2} \ln \left( \frac{L}{H } \right)$, are recovered for $H \to L \gg \lD$ and for $H \to \lD \ll L$, respectively.

In quasi-2D systems, $L \gg H \gg \lambda_D$, Eq.~\ref{eq:dw} is approximated by
\begin{equation}
{\rm DW} \simeq {\rm DW_{3D}} \left[1+\frac{\lD}{H}\left( \ln \left(\frac{L}{H}\right)-1\right)\right].
\label{eq:dwapp}
\end{equation}
Hence, we predict that in confined systems the DW grows logarithmically with $L$, as in 2D, with a slope decreasing as $1/H$. 
We remark here that, as long as $H \gg \lD$, the DW factor {\it grows} as $H$ decreases at constant $L$, e.g., as the system becomes more confined.
This occurs because, as $H$ decreases, a larger fraction of the phonon spectrum becomes effectively two-dimensional.

\section{Numerical details}
We validate our theoretical prediction, and explore the effect of confinement on the liquid phase, via extensive molecular dynamics simulations~\cite{plimpton1995fast} of the standard A:B 80:20 Kob-Andersen (KA) Lennard-Jones (LJ) mixture \cite{kob1994scaling}, where particles interact via the potential 
$V_{\alpha \beta}\left({r}\right) = 4\epsilon_{\alpha \beta}\left[\left(\frac{\sigma_{\alpha \beta}}{r}\right)^{12}-\left(\frac {\sigma_{\alpha \beta}} {r}\right)^{6} + {C}_ {\alpha \beta}\right]$, and $\epsilon_{AB} = 1.5\epsilon_{AA}$, $\epsilon_{BB} = 0.5\epsilon_{AA}$, $\sigma_{AB} = 0.8\sigma_{AA}$, $\sigma_{BB} = 0.88\sigma_{AA}$, $\alpha, \beta \in \left\{ {A,B} \right\}$.
The potential is truncated at $r_c = 2.5\sigma_{\alpha \beta}$, and $C_{\alpha\beta}$ enforces $V(r_c) = 0$.
The mass of the particles $m$, $\epsilon_{AA}$, and $\sigma_{AA}$ are our unit of mass, energy and distance, respectively.
We first thermalize the system in the NPT ensemble, at $P = 1.0$, allowing the box size to vary only in the lateral dimensions. 
Production runs are then performed in the NVE ensemble.
The number of particles depends on $L$ and $H$, and varies between $10^3$ to $~10^6$ million.
We average the dynamical data over at least four independent runs.

We monitor the relaxation dynamics studying the mean square displacement,
$\langle \Delta r^2(t)\rangle = \frac{1}{N}\sum \Delta \mathbf{r}_i^2(t)$, where $\Delta \mathbf{r}_i$ is the displacement of particle $i$ at time $t$, and the self-scattering function, 
$F_s\left(k,t \right) = \frac{1}{N} \left< \sum_{j=1}^N e^{i\mathbf{k}\cdot\Delta\mathbf{r}_j \left(t\right)}\right>$,
where $\mathbf{k}$ the wavevector of the first peak of the static structure factor of bulk systems.
The relaxation time $\tau$ is defined by $F_s\left(k,\tau \right) = 1/e$.
We further investigate the dynamics using the cage-relative mean square displacement and self-scattering function~\cite{shiba2016unveiling, illing2017mermin, vivek2017long,Tong2018}.
These are defined as above, with the displacement of particle $i$ replaced by its cage-relative counterpart, 
$\Delta_{\rm CR} \mathbf{r}_i = \Delta \mathbf{r}_i - \frac{1}{n_i}\sum_j\Delta \mathbf{r}_j$, where the sum is over all neighbors of particle $i$ at time $t = 0$. We identify the neighbors via the Voronoi construction.
\begin{figure}[!!t]
\centering
\includegraphics[angle=0,width=0.95\columnwidth]{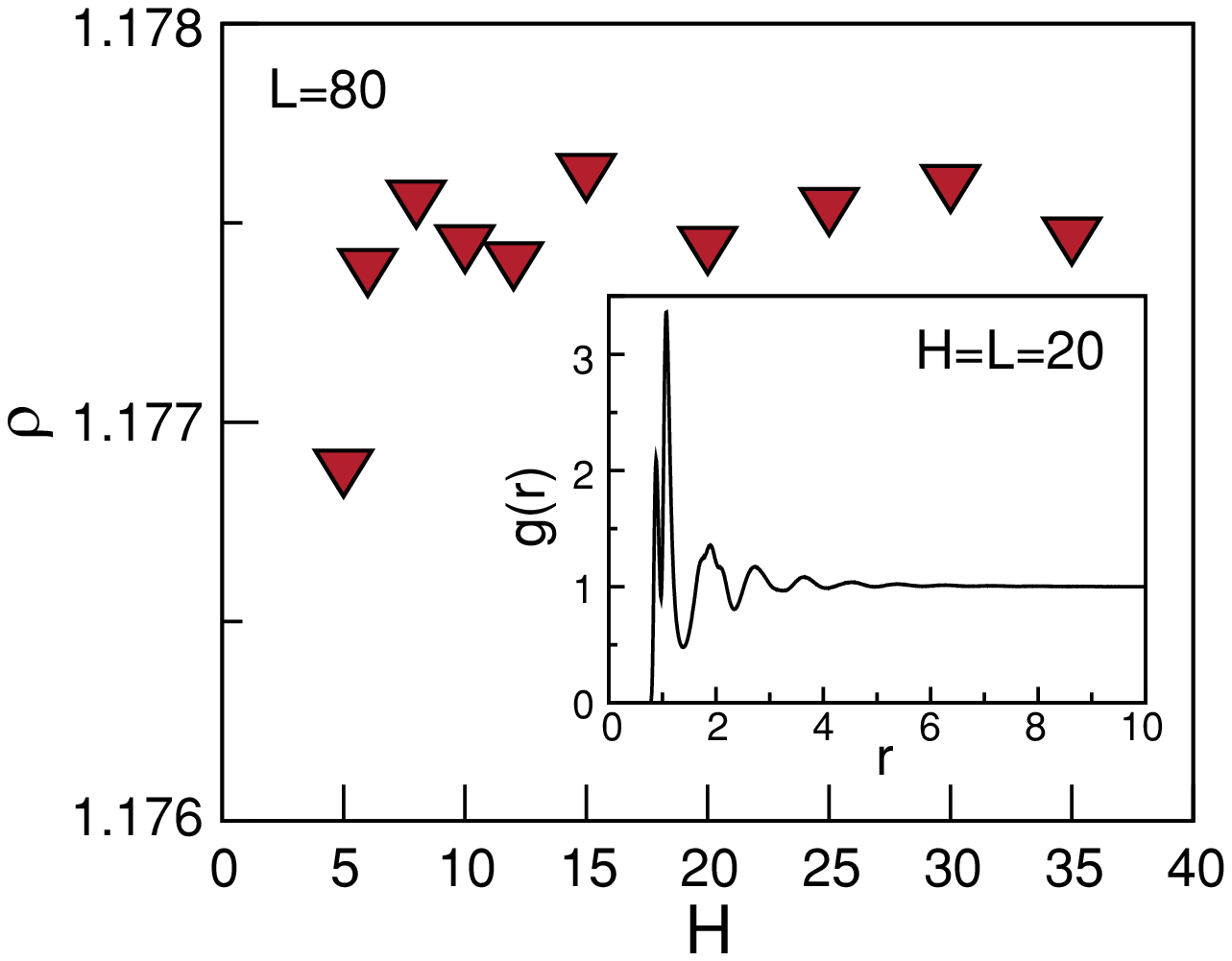}
\caption{
The dependence of the average density on the gap width, at $T = 0.35$, when periodic boundary conditions are used in the confining direction. 
Confinement does not strongly influence the average density, in the range of gap widths we have considered. 
The radial distribution function, shown in the inset for $H = 20$ and $L = 20$ at $T = 0.35$, approaches one at $\xib/2 \simeq 5$.
\label{fig:density}
}
\end{figure}

We consider three different confinement approaches.
First, we use periodic boundary conditions in the confining direction, which is an approach that is useful to avoid layering as well as to compare with the theoretical predictions. 
When using this approach, the density is essentially constant, $\rho = 1.1775(5)$, as we illustrate in Fig.~\ref{fig:density}(a). 
In the figure we also show that, for larger $H$ values representative of the bulk limit, the radial distribution function becomes constant for $r \simeq \xib/2 \simeq 5$.

Secondly, we confine the system between flat walls.
In this case, the interaction between particles of type $i=A,B$ and the walls is given by a LJ potential with energy scale $\epsilon_{ii}$ and length scale $\sigma_{ii}$, truncated in its minimum.
In the presence of flat walls, the density sensibly decreases with $H$, and layering occurs, as shown in Fig.~\ref{fig:density}b.

Finally, we perform simulations of systems confined between rough walls.
In this case, we first thermalize at the desired state pressure large samples, using periodic boundary conditions in all directions, and then freeze the positions of all particles whose height is outside the interval $[0:H]$. 
When using rough walls, we work at fixed density rather than at fixed pressure.

\section{Confined amorphous solids}
We study the density of states of confined amorphous solid configurations generated by minimizing the energy of configurations equilibrated at low temperature.
We fix the pressure of these low-temperature configurations to $P =  1$ by adjusting the lateral size, which slightly fluctuates around $L = 80$. 
We considered several $H$ values, so that the number of particles ranges from $36000$ to $150000$.
We further use periodic boundary conditions in all spatial directions to prevent structural inhomogeneities due to layering, hence allowing for a more transparent comparison with the theoretical predictions. 
The effect of walls is discussed in Sec.~\ref{sec:walls}.
\begin{figure*}[!t]
\centering
\includegraphics[angle=0,width=2\columnwidth]{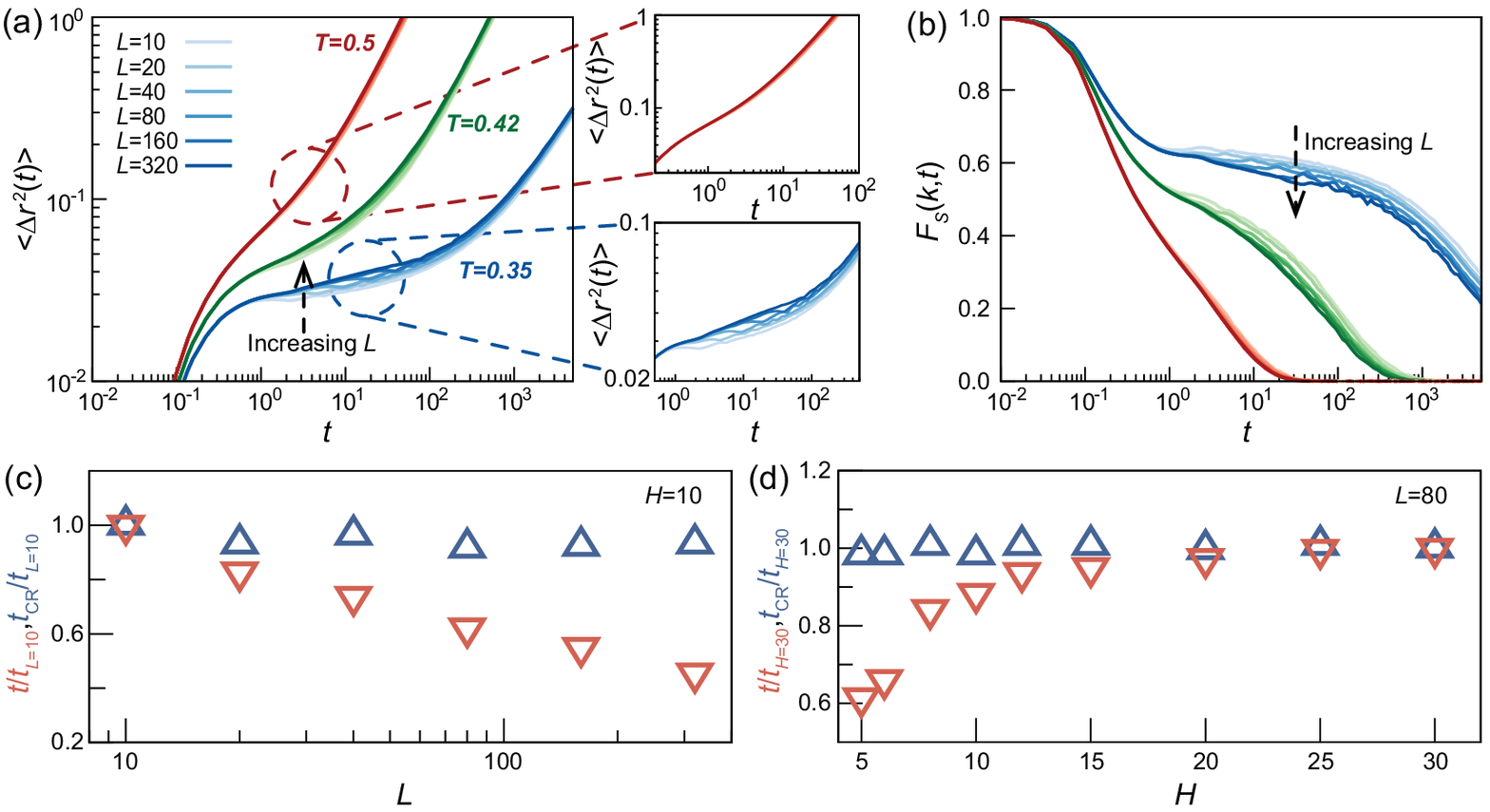}
\caption{ Long-wavelength fluctuations in confined amorphous solids. (a) Mean square displacement, and (b), self-scattering function, at three different values of the temperature.
We fix $H = 10$ and show, at each temperature, results for $10 \leq L \leq 320$. 
(c) The relaxation time decreases as the lateral size increases, while the cage-relative relaxation time is $L$-independent.
(d) The relaxation time decreases as the gap-size decreases, particularly for $H \leq \xib$, while the cage-relative relaxation time is $H$-independent.
The relaxation times in (c) and (d) are divided by their respective values at $L = 10$ and at $H = 30$, to facilitate their comparison.
In (c) and (d), errors are smaller than the symbol size.
\label{fig3:diffT}
}
\end{figure*}

We evaluate the low-frequency end of the vibrational spectrum of the generated energy minima via the direct diagonalization of their Hessian matrix. 
To compare the numerical results with our theoretical prediction of Eq.~\ref{eq:Domega}, schematically illustrated in Fig.~\ref{fig:dos}(a), we focus on the frequency dependence of the cumulative distribution $C(\omega) = \int D(\omega) d\omega$.
Due to the large lateral size of our systems~\cite{Tanguy2002}, we observe gaps at low frequency, as predicted by linear elasticity~\footnote{We have verified that these gaps are not an artefact of the discontinuity of the force at the cutoff distance~\cite{Shimada2018a}, as they persist when the interaction potential is appropriately smoothed.}.
Figire~\ref{fig:dos}(b) also demonstrates that $C(\omega)/\omega^2$ is constant at small frequencies, and increases above an $H$ dependent crossover frequency which, according to Eq.~\ref{eq:Domega}, should scale as $\omega_H \propto c_s/H$.
Indeed, when plotted versus $\omega H$, and vertically scaled, the data collapse up to their crossover point, as we illustrate in Fig.~\ref{fig:dos}(c).
The figure also supports the $\omega^2$ to $\omega^3$ crossover for the cumulative distribution suggested by the theoretical model.

We remark that the data collapse of Fig.~\ref{fig:dos}(c) breaks for small $H$.
To rationalize this observation, we investigate in Fig.~\ref{fig:density} the gap size dependence of the density and the radial correlation function of a low-temperature solid configuration. 
We observe that the density is almost $H$ independent, for $H \geq 5$, and that the radial correlation function approaches the ideal gas limit at $r \simeq 5$.
This allows us to estimate the structural correlation length of the bulk solid, $\xi_{\rm bulk} \simeq 10$.
We thus understand that, in Fig.~\ref{fig:dos}(c), no collapse occurs for small $H$ as confinement interferes with the structural correlation length of the system.

We further validate our theoretical prediction for the dependence of the DW factor of amorphous solids on the relevant length scales $L$ and $H$, Eq.~\ref{eq:dwapp}, performing simulations at a low-temperature value at which structural relaxation is negligible.
In this limit, the mean-square displacement approaches a constant DW value at long times, as illustrated in Fig.~\ref{fig:dos}(d) for $H = 10$.
Figure~\ref{fig:dos}(e) shows that this limiting DW factor grows as the logarithm of the lateral size $L$, with a slope scaling as $1/H$, in agreement with the predictions of Eq.~\ref{eq:dwapp}.

\begin{figure*}[!t]
\centering
\includegraphics[angle=0,width=2\columnwidth]{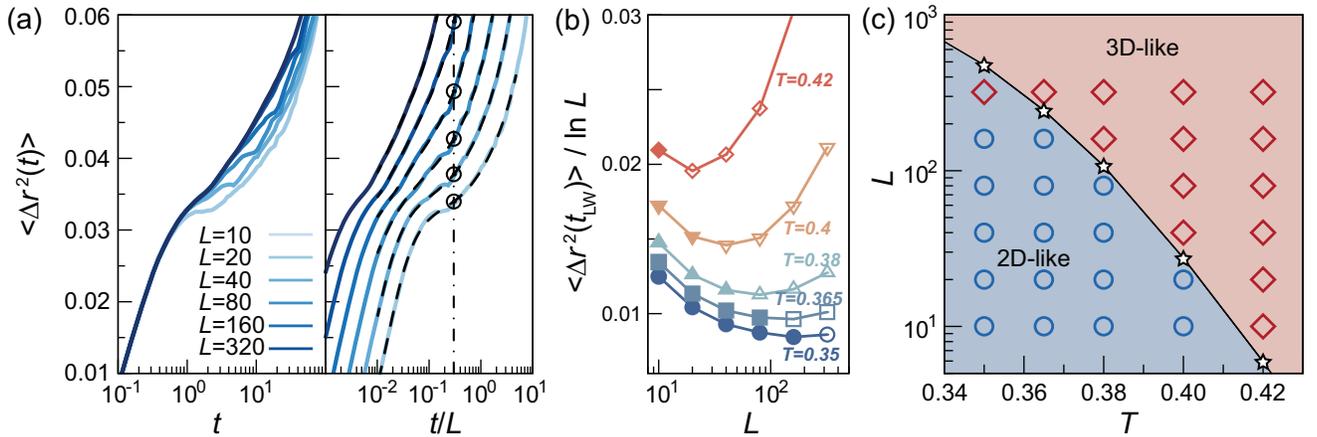}
\caption{
Dimensionality crossover in confined liquids. (a) The mean square displacement exhibits a crossover between two different regimes at a time $t_{\rm LW} \simeq 0.3 L$ (dash-dotted line). 
Dashed lines are polynomial fits used to estimate the mean square displacement at the crossover time (circles).
Data are for $T = 0.35$, and different $L$ values.
(b) The mean square displacement at the crossover time grows faster that $\ln L$ (open symbols), above a characteristic $T$ dependent lateral system size. 
When this occurs, structural relaxation rather than LWs dominate the diffusivity, and hence the system has a 3D-like behaviour.
(c) State points with an effective two-dimensional behaviour according to the analysis in (b), are illustrated as open circles. 
Diamonds, conversely, identify those having a three-dimensional behaviour. 
Stars correspond to the prediction of Eq.~\ref{eq:lstar}, $L = \alpha c_s \tau_{CR}(T)$, with $\alpha \simeq 0.018$. 
The interpolating solid line is a guide to the eye. 
All panels refer to $H = 10$. 
Supplemental Fig. S4 shows that the results are insensitive to changes in the gap width.
\label{fig4:diffW0.35T}}
\end{figure*}

\section{Confined liquids} 
Having ascertained that LWs influence the behaviour of confined solids, we now demonstrate that they similarly affect the relaxation dynamics of quasi-2D supercooled-liquids.
To this end, we investigate the size and temperature dependence of the mean square displacement and self-scattering function at the wave vector of the peak of the static structure factor of bulk systems.
Figures~\ref{fig3:diffT}(a) and (b) show that the transient solid-like response revealed by the mean square displacement and the self-scattering function becomes less apparent as the system size decreases.
This size dependence is more apparent at low temperature, where the transient solid like behaviour is manifest.

We prove that this observed size dependence originates from LW fluctuations by comparing the $L$ dependence of the relaxation time $\tau$ and of the cage-relative (CR) relaxation time $\tau_{\rm CR}$.
Cage-relative quantities, indeed, are insensitive to collective particle displacements and hence filter out the effect of LWs ~\cite{shiba2016unveiling, illing2017mermin, vivek2017long}. 
In Fig.~\ref{fig3:diffT}(c), we observe that, while the standard relaxation time decreases logarithmically with $L$, the CR one is $L$ independent.
These results closely parallel those observed in strictly two-dimensional systems~\cite{flenner2015fundamental, shiba2016unveiling, illing2017mermin, vivek2017long, zhang2019long, shiba2019local,li2019long} and demonstrate that LW fluctuations sensibly affect the structural relaxation dynamics of confined liquids.

In Fig.~\ref{fig3:diffT}(d), we further show that the relaxation time $\tau$ decreases as the gap width is reduced and a larger fraction of the vibrational spectrum becomes effectively two-dimensional. 
This dynamical speed up is particularly relevant for $H < \xib$, indicating that the structural changes induced by such strong confinement promote LW fluctuations.
This is consistent with the observation of a significant increment in the density of low-frequency modes for $H = 5$, in Fig.~\ref{fig:dos}(b).
The gap independence of the cage-relative relaxation time, also illustrated in Fig.~\ref{fig3:diffT}(d), confirms our interpretation, namely that the $H$-induced speed-up originates from LW fluctuations.
\begin{figure*}[!!t]
\centering
\includegraphics[angle=0,width=2\columnwidth]{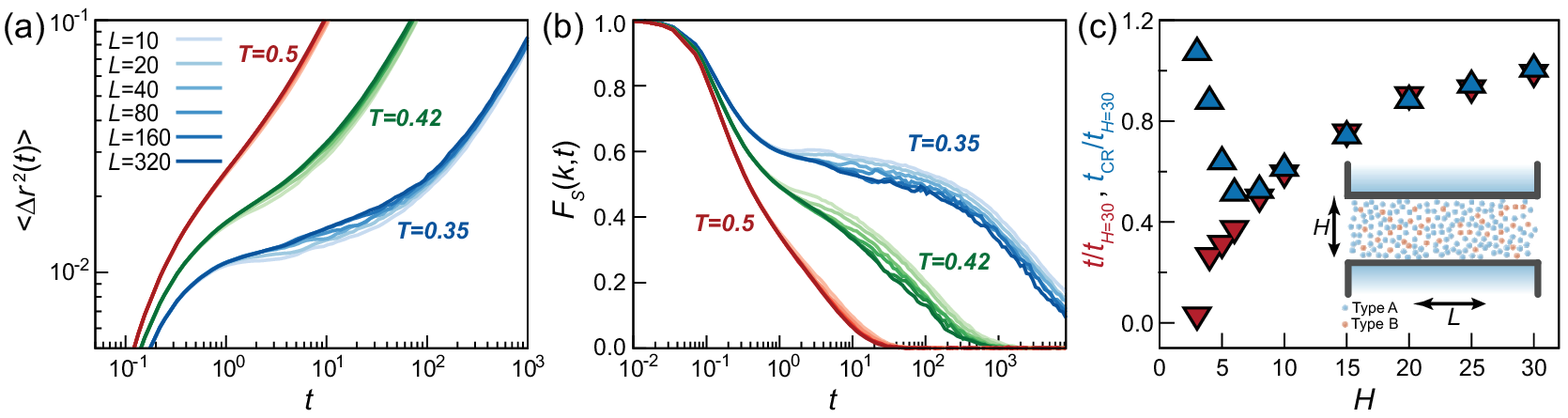}
\caption{
Long-wavelength fluctuations in slit geometries. (a) The transverse mean-square-displacement and (b) the self-intermediate scattering function for supercooled liquids with various transverse length scales $10 \leq L \leq 320$ at the same perpendicular length scale $H = 10$. (c) Width dependence of the relaxation time, and of the cage-relative relaxation time for a $L = 40$ system.
Errors are smaller than the symbol size.
The inset is a schematic diagram of the confining geometry.
\label{fig:walls}
}
\end{figure*}

We quantitatively investigate the dimensionality crossover focusing on the mean square displacement, $\langle \Delta r^2(t) \rangle$.
In the solid phase, $\langle \Delta r^2(t) \rangle$ approaches an asymptotic DW factor value on a time scale $t_{\rm LW} \propto \omega_L^{-1}\propto L$.
The asymptotic value of the DW factor grows as $\ln L/f(L)$, with $f(L)$ a slowing increasing function of $L$, corresponding to the denominator of Eq.~\ref{eq:dw}.
In the liquid phase, therefore, we expect a crossover in the time dependence of the mean square displacement at a time $t_{\rm LW}$.
Figures~\ref{fig4:diffW0.35T}(a) and \ref{fig4:diffW0.35T}(b) demonstrate that such a crossover occurs at $t_{\rm LW} \simeq 0.3 L$, for $T = 0.38$. 
At the same $t_{\rm LW}$ similar crossovers occur at all temperatures.

When LW fluctuations dominate the dynamics, as in the solid phase, $\frac{\langle \Delta r^2(t_{\rm LW})}{\ln L} \rangle \propto 1/f(L)$ decreases with $L$.
We therefore assume LW fluctuations to become negligible at $L$ values at which $\langle \Delta r^2(t_{\rm LW}) \rangle$ grows faster than $\ln L$. 
When this occurs, irreversible relaxation events rather than large-amplitude oscillations dominate the diffusivity.
In Fig.~\ref{fig4:diffW0.35T}(b) we indeed observe that $\langle \Delta r^2(t_{\rm LW}) \rangle/\ln L$ is not monotonic in $L$, decreasing with $L$ when LWs are relevant (solid symbols), and increasing when they are not (open symbols). 
This behavior allows us to identify crossover $L$ values, which we have verified not to depend on the gap width.
This study leads to the $L$-$T$ diagram of Fig.~\ref{fig4:diffW0.35T}(c).
The system-size dependent dynamics characteristic of two-dimensional behaviour occurs at low temperature and small lateral size and disappears as either the lateral length or the temperature increase.
We remark that while this diagram does not depend on the confinement width $H$, size effects gradually fade away as $1/H$, as in the solid phase, and hence become not appreciable at large $H$.

We exploit the size independence of the cage-relative relaxation time to rationalize this observed dimensionality crossover.
Indeed, vibrational excitations cannot last more than the cage-relative relaxation time, as on this time scale the structure of the system sensibly changes, as particles change neighbours. 
Since the vibrational modes influencing the structural relaxation dynamics are those that have time to develop, we expect the crossover between a two-dimensional size-dependent relaxation dynamics and a three-dimensional size-independent relaxation dynamics to occur at
\begin{equation}
\frac{L}{\tau_{\rm CR}(T)} = \alpha c_s,
\label{eq:lstar}
\end{equation}
with $c_s$ being the transverse sound velocity and $\alpha$ being a constant.
In other words, size effects disappear for $L > \alpha c_s \tau_{\rm CR}(T)$, as the system relaxes before the lowest size-dependent mode develops. 
This theoretical prediction well describes the data of Fig.~\ref{fig4:diffW0.35T}(d).

\section{Effect of smooth and rough walls~\label{sec:walls}}
Our theoretical analysis and numerical simulations demonstrate that LW fluctuations affect the dynamics of confined liquids.
However, so far we have described simulations obtained using periodic boundary conditions in all spatial directions;
one might wonder, therefore, whether LWs also play a role in the experimentally relevant set up of liquids confined between two parallel walls at a separation $H$.
To address this question, we investigate the relaxation dynamics of the KA LJ binary mixture confined  between two atomically-smooth flat walls.
Since the walls prevent diffusion along the transverse direction, we focus on particle motion in the lateral directions, effectively defining two-dimensional mean-square displacement and self-scattering function.
We find that, under wall confinement, the relaxation dynamics has the typical size dependence induced by LW fluctuations, the caging regime becoming less apparent as $L$ increases, as we illustrate in Figs.~\ref{fig:walls}(a) and \ref{fig:walls}(b).

\begin{figure}[!!t]
\centering
\includegraphics[angle=0,width=0.95\columnwidth]{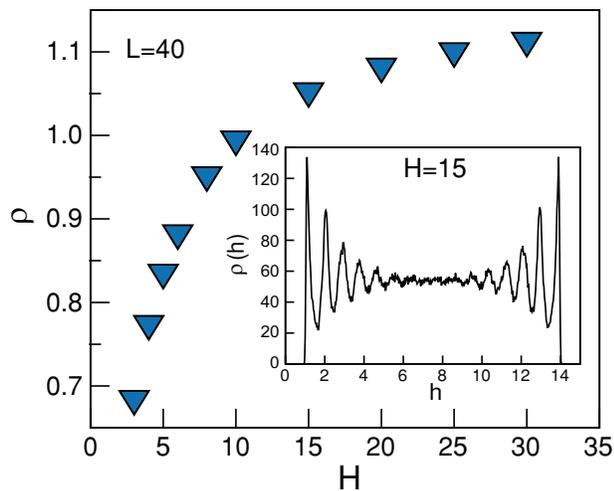}
\caption{
Dependence of the average density on the gap width, at $T = 0.35$, for a system confined in between flat walls at a separation $H$.
The density decreases as the gap width decreases.
Flat walls, furthermore, induce layering, as we illustrate in the inset by plotting the density at a distance $h$ from a confining wall.
\label{fig:densitywall}
}
\end{figure}

\begin{figure}[!!t]
\centering
\includegraphics[angle=0,width=0.92\columnwidth]{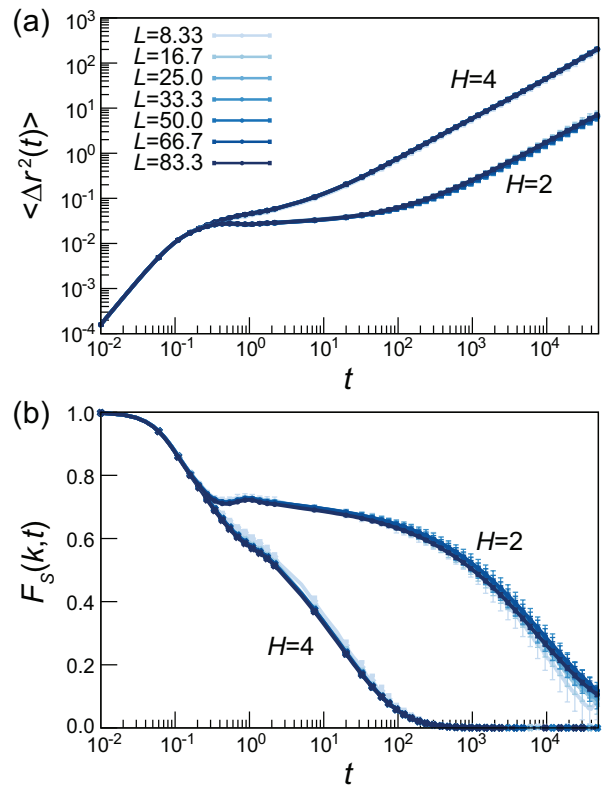}
\caption{
Mean-square displacement (a) and self-scattering function (b) of systems confined between rough walls, at $H=2$ and $H = 4$. 
These quantities are evaluated focusing on the behavior of the central layer of particles.
The relaxation dynamics does not depend on the lateral length, or system size $N$, indicating that the rough walls kill the LW fluctuations.
\label{fig:msdRoughwall}
}
\end{figure}

The structural changes induced by the walls, however, strongly affect the relaxation dynamics, as evidenced by the $H$ dependence of the standard and CR relaxation times, which we illustrate in Fig.~\ref{fig:walls}(c).
For $H \geq \xib$, both relaxation times decrease as the system becomes more confined; this is, we believe, the combined effect of layering and the reduction in the average density induced by the confinement, which we illustrate in Fig.~\ref{fig:densitywall}.

Importantly, we observe in Fig.~\ref{fig:walls}(c) that for $H \leq \xib$, while the relaxation time decreases as the gap width is reduced, the cage-relative relaxation sharply increases.
This increase in the CR-relaxation time is in qualitative agreement with the many previous investigations reporting an increase in the viscosity of molecular liquids under confinement~\cite{Granick1991, Hu1991, Demirel1996, Demirel2001, Kienle2016}. 
Indeed, we remind the reader that viscosity and cage-relative relaxation time are related~\cite{Flenner2019,li2019long}.
This observed decoupling demonstrates that smooth walls do not kill the LWs, but rather make their effect more apparent.

While smooth walls do not kill LWs, rough walls strongly suppress them.
Indeed, we show in Fig.~\ref{fig:msdRoughwall} that the relaxation dynamics of liquids confined between rough walls does not depend on the lateral system size.
We remark that for very large gap widths the effect of the boundary should become negligible, and hence LW fluctuations should play a role.
Since the influence of LWs on the dynamics scales as $1/H$, however, their effect in this large-$H$ limit may be not easily appreciated. 
We expect variations~\cite{jabbarzadeh2006low} in the roughness of the confining walls and the wall-liquid interaction potential to only  qualitatively affect the observed phenomenology.

\section{Conclusions and experimental relevance}
The confinement-induced enhancement of the DW factor described Eq.~\ref{eq:dw} is an equilibrium property not affected by the underlying microscopic dynamics, equally valid for molecular and colloidal solids.
In the supercooled regime, the signatures of LW fluctuations conversely depend on how much the system moves along the phase-space directions of the low-frequency modes before particles rearrange.
Since the size of this displacement depends on the microscopic dynamics and it is smaller if the system moves diffusively, rather than ballistically, we expect the influence of confinement to be more relevant at the molecular scale rather than at the colloidal scale.
Nevertheless, we remind the reader that LWs are observed in experiments~\cite{vivek2017long,li2019long,zhang2019long} and simulations~\cite{li2019long} of two-dimensional colloidal systems;
our predictions concerning the role of LW fluctuations in confined systems therefore apply to both molecular and colloidal systems.

For the effect of LW fluctuations to be experimentally visible, however, the roughness scale of the confining walls must be smaller than the size of the particles. 
Rough walls, indeed, affect the motion in the lateral dimensions and kill the LW fluctuations, as we have shown in Fig.~\ref{fig:msdRoughwall}.
The requirement of smooth confining walls is not a technical limitation. 
Walls that are {\it de facto} flat at the molecular scale exist~\cite{Zhu2004}, and it is undoubtedly possible to confine large colloidal particles between walls that are flat at the particle scale.
In colloidal experiments, however, one should ascertain that no particles stick irreversibly to the walls, effectively making them rough, e.g., as observed in Refs.~\cite{Nugent2007, Edmond2012}.
Hence our predictions are experimentally testable both in confined molecular liquids, e.g., comparing the size dependence of the viscosity and of structural relaxation time, and in confined colloidal systems, comparing, e.g. the standard and cage-relative relaxation times.

Our results show that confined systems exhibit a gradual dimensionality crossover controlled by the gap width and the temperature, which is appreciable when investigating the lateral size dependence of the dynamics. 
The physics of confined liquids is thus richer than previously realised. 
These findings might be relevant to a variety of applications involving micro- and nanofluidics, e.g., lab-on-a-chip devices, where particles flow in confined geometries.

\begin{acknowledgments}
We acknowledge support from the Singapore Ministry of Education through the Academic Research Fund Tier RG 86/19(S), Singapore, and the National Supercomputing Centre Singapore (NSCC) for the computational resources. 
\end{acknowledgments}


%
\end{document}